\documentclass[aps,prb,twocolumn,superscriptaddress]{revtex4} 
\usepackage[dvips]{graphicx}
\usepackage{bm}% bold math
\usepackage{latexsym} 
\usepackage{amsmath}
\usepackage{amssymb}

\newcommand{\ui}{{\rm i}} 

\newcommand{\bmr}{{\bm r}}

\newcommand{\bms}{{\bm s}}

\newcommand{\bmq}{{\bm q}}

\newcommand{\bmk}{{\bm k}}

\newcommand{\bmp}{{\bm p}}

\newcommand{\kB}{k_{\rm B}}

\begin{document}

\title{Linear-response theory of spin Seebeck effect 
in ferromagnetic insulators}

\author{
Hiroto Adachi
}
\email[]{hiroto.adachi@gmail.com}
\affiliation{
Advanced Science Research Center, Japan Atomic Energy Agency, 
Tokai 319-1195, Japan
} 
\affiliation{
CREST, Japan Science and Technology Agency, Sanbancho, Tokyo 102-0075, Japan
}
\author{
Jun-ichiro Ohe
} 
\affiliation{
Advanced Science Research Center, Japan Atomic Energy Agency, 
Tokai 319-1195, Japan
}
\affiliation{
CREST, Japan Science and Technology Agency, Sanbancho, Tokyo 102-0075, Japan
}
\author{
Saburo Takahashi
}
\affiliation{
Institute for Materials Research, 
Tohoku University, Sendai 980-8577, Japan
}
\affiliation{
CREST, Japan Science and Technology Agency, Sanbancho, Tokyo 102-0075, Japan
}
\affiliation{
Advanced Science Research Center, Japan Atomic Energy Agency, 
Tokai 319-1195, Japan
}
\author{
Sadamichi Maekawa
}
\affiliation{
Advanced Science Research Center, Japan Atomic Energy Agency, 
Tokai 319-1195, Japan
}
\affiliation{
CREST, Japan Science and Technology Agency, Sanbancho, Tokyo 102-0075, Japan
}

\date{\today}

\begin{abstract} 
We formulate a linear response theory of the spin Seebeck effect, i.e., 
a spin voltage generation from heat current flowing in a ferromagnet. 
Our approach focuses on the collective magnetic excitation of 
spins, i.e., magnons. 
We show that the linear-response formulation provides us with 
a qualitative as well as quantitative understanding of 
the spin Seebeck effect observed in a prototypical magnet, 
yttrium iron garnet. 

\end{abstract} 



\maketitle 

\section{Introduction \label{Ch:intro}}

The generation of spin voltage, i.e., the potential for an electron's spin  
to drive spin currents, by a temperature gradient in a ferromagnet 
is referred to as the spin Seebeck effect (SSE). 
Since the first observation of the SSE in a 
ferromagnetic metal Ni$_{81}$Fe$_{19}$,~\cite{Uchida08} 
this phenomenon has attracted much attention 
as a new method of generating spin currents from heat energy 
and opened a new possibility of spintronics devices.~\cite{Zutic04} 
The SSE triggered the emergence of the new field dubbed 
``spin caloritronics''~\cite{SpinCalo,Slonczewski} in the 
rapidly growing spintronics community. 
Moreover, as the induced spin voltage can be converted into electric 
voltage through the inverse spin Hall effect~\cite{Saitoh06} 
at the attached nonmagnetic metal, this phenomenon put a new twist 
on the long and well-studied history of thermoelectric research.~\cite{Blatt76} 

One of the canonical frameworks to describe nonequilibrium transport 
phenomena is linear-response theory.~\cite{Mahan-text} 
Having been applied to a number of transport phenomena, 
linear-response theory has been so successful 
because it is intimately related 
to the universal fluctuation-dissipation theorem. 
Up to now, however, the linear-response formulation of the SSE has not 
been known mainly because, unlike the charge current, 
the spin current is not a conserved quantity. 
Therefore, it is of great importance to formulate the SSE in terms 
of linear-response theory. 

Concerning the SSE, a big mystery is now being established, which is, 
how can conduction electrons sustain the spin voltage over 
such a long range of several millimeters~\cite{Uchida08} 
in spite of the conduction electrons' short spin-flip 
diffusion length, which is typically of several tens of nanometers? 
A key to resolve this puzzle was reported by 
a recent experiment on electric signal transmission through 
a ferromagnetic insulator~\cite{Kajiwara10} 
which demonstrates that the spin current can be carried by 
the low-lying magnetic excitation of {\it localized} spins, i.e., 
the magnon excitations, and that it can transmit the spin current 
as far as several millimeters. 
Subsequently, the SSE was reported to be observed in the  
magnetic insulator LaY$_2$Fe$_5$O$_{12}$ despite the absence of 
conduction electrons.~\cite{Uchida10} These experiments suggest that 
contrary to the conventional wisdom over the last two decades that 
the spin current is carried by {\it conduction} electrons,~\cite{Maekawa06} 
the magnon is a promising candidate as a carrier for the SSE. 

The purpose of this paper is twofold. First, we analyze the SSE 
observed in LaY$_2$Fe$_5$O$_{12}$~\cite{Uchida10} 
(hereafter referred to as YIG) in terms of magnon spin current, 
i.e., a spin current carried by magnon excitations. 
Second, we develop a framework for analyzing the SSE 
by means of the standard linear-response formalism~\cite{Mahan-text} 
which is amenable to the language of the magnetism community.~\cite{Simanek03} 
This allows us to describe the spin transport phenomena 
systematically, and it can be easily generalized 
to a situation including degrees of freedom other than magnons, 
e.g., conduction electrons and phonons, 
to describe a more complicated process in the case of 
metallic systems.~\cite{Uchida08}

The plan of this paper is as follows. 
In Sec.~\ref{Sec:local}, we present a linear-response 
approach to the ``local'' spin injection by thermal magnons, 
in which the spin injection is driven by the temperature 
difference between the ferromagnet and the attached nonmagnetic metal. 
Next, in Sec.~\ref{Sec:nonlocal} we develop a linear-response 
theory of the ``nonlocal'' spin injection by thermal magnons, 
in which the spin injection is driven by the temperature 
gradient inside the ferromagnet. 
As one can see below, this process can explain the SSE observed 
in YIG.~\cite{Uchida10} 
Finally, in Sec.~\ref{Sec:conclusion} we summarize and discuss our results.

\section{``Local'' spin injection by thermal magnons \label{Sec:local}} 

%%%%%%%%%%%%%%%%%%%%%%%%%%%%%%%%%%%%%
\begin{figure}[t] 
  \begin{center}
    \scalebox{0.3}[0.3]{\includegraphics{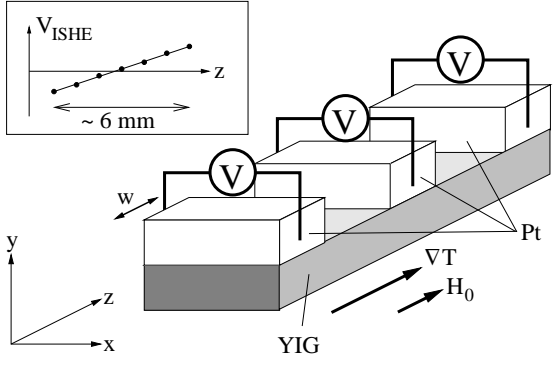}}
  \end{center}
\caption{Experimental setup for observing the SSE.~\cite{Uchida10} 
Inset: Schematics of the spatial profile of the observed voltage. 
}
\label{Fig:fig1_SSEins}
\end{figure} 
%%%%%%%%%%%%%%%%%%%%%%%%%%%%%%%%%%%%%

We start by briefly reviewing the SSE experiment for YIG.~\cite{Uchida10} 
Figure~\ref{Fig:fig1_SSEins} shows the experimental setup  
where several Pt terminals are attached on top of a YIG film in 
a static magnetic field $H_0 \hat{\bm z}$ ($\gg$ anisotropy field) 
which aligns the localized magnetic moment along $\hat{\bm z}$. 
A temperature gradient ${\bm \nabla}T $ 
is applied along the $z$ axis, and it 
induces a spin voltage across the YIG/Pt interface. 
This spin voltage then injects a spin current $I_s$ into the Pt terminal 
(or ejects it from the Pt terminal). 
A part of the injected/ejected spin current $I_s$ is converted 
into a charge voltage 
through the so-called inverse spin Hall effect:~\cite{Saitoh06} 
%%%
\begin{equation}
V_{\rm ISHE}= \Theta_H (|e|I_s)(\rho/w), 
\label{Eq:ISHE01}
\end{equation}
%%%
where $|e|$, $\Theta_H$, $\rho$, and $w$ are 
the absolute value of electron charge, spin Hall angle, resistivity, 
and width of the Pt terminal (see Fig.~\ref{Fig:fig1_SSEins}), 
respectively. Hence, the observed charge voltage $V_{\rm ISHE}$ 
is a measure of the injected/ejected spin current $I_s$.

\begin{figure}[t] 
  \begin{center}
    \scalebox{0.65}[0.65]{\includegraphics{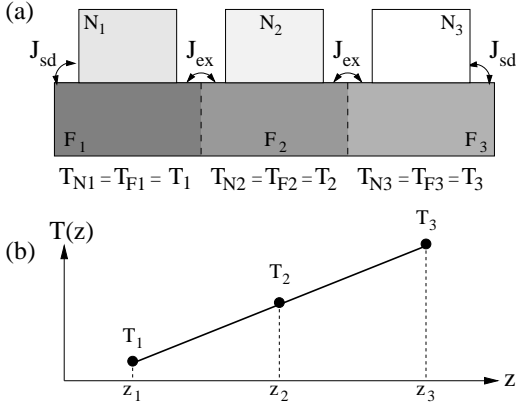}}
  \end{center}
\caption{(a) System composed of ferromagnet 
($F$) and nonmagnetic metals ($N$) divided into the three temperature 
domains of $F_1/N_1$, $F_2/N_2$, and $F_3/N_3$ with their local 
temperatures of $T_1$, $T_2$, and $T_3$. 
(b) Temperature profile.} 
\label{Fig:fig2_SSEins}
\end{figure}

To investigate the SSE observed in YIG, we consider a model 
shown in Fig.~\ref{Fig:fig2_SSEins}(a). 
While YIG is a ferrimagnet, we model it as a ferromagnet 
since we are interested in the low-energy properties. 
The key point in our model is that the temperature gradient 
is applied over the insulating ferromagnet, but there is {\it locally} 
no temperature difference between the ferromagnet and 
the attached nonmagnetic metals, i.e., $T_{N_1}= T_{F_1}= T_1$, 
$T_{N_2}= T_{F_2}= T_2$, and $T_{N_3}= T_{F_3}=T_3$. 
We assume that each domain is initially in thermal equilibrium 
without interactions with the neighboring domains, and then calculate 
the nonequilibrium dynamics after we switch on the interactions. 
Note that this procedure is essentially equivalent to that 
used by Luttinger~\cite{Luttinger64} to realize the initial condition 
mentioned above.

Let us consider first the low-energy excitations 
in the ferromagnet. In the following, we focus on the spin-wave region 
where the magnetization ${\bm M}(\bmr)$ 
fluctuates only weakly around the ground state value $M_s \hat{\bm z}$ 
with the saturation magnetization $M_s$, 
and we set 
${\bm M}/M_s= (1- m^2/2) \hat{\bm z}+ {\bm m}$ 
to separate the small fluctuation part ${\bm m}$ $(\perp \hat{\bm z})$ 
from the ground state value. 
Then, the low-energy excitations of ${\bm M}$ are described by 
boson (magnon) operators $a_\bmq^\dag$ and $a_\bmq$ 
through the relations~\cite{Akhiezer60} $m^+_\bmq = \sqrt{1/S_0} a^\dag_{-{\bm q}}$ 
and $m^-_\bmq = \sqrt{1/S_0} a_{\bm q}$ 
where $m^\pm \equiv (m^x \pm \ui m^y)/\sqrt{2}$, $S_0$ is the 
size of localized spins, and 
${\bm m}({\bm r},t)=  
{N_F}^{-1/2} \sum_{\bm q} {\bm m}_{\bm q}(t) e^{\ui {\bm q}\cdot {\bm r}}$ 
with $N_F$ being the number of localized spins in the ferromagnet. 
Consistent with this boson mapping, the magnetization dynamics 
is described by the following action:~\cite{Doniach74,Schmid82} 
%%%
\begin{eqnarray}
  {\cal S}_F &=&  \int_{C} dt \; \sum_{\bmq} 
  m^+_{-\bmq}(t) [X_\bmq(\ui \partial_t)]^{-1} m^-_{\bmq}(t), 
  \label{Eq:S_F01} 
\end{eqnarray}
%%%
where the integration is performed along the Keldysh contour $C$,~\cite{Rammer86} 
and the bare magnon propagator is given by 
%%%
\begin{eqnarray}
\check{X}_{\bmq}(\omega) &=&
\left( { X^R_\bmq (\omega), \atop 0,} 
     { X^K_\bmq (\omega)\atop X^A_\bmq (\omega)} \right) 
\end{eqnarray}
%%% 
with the following equilibrium condition: 
%%%
\begin{eqnarray}
X^A_\bmq(\omega)= [X^R_\bmq(\omega)]^*,&&
X^K_\bmq(\omega)= 2 \ui \, {\rm Im} X^R_\bmq(\omega) 
\coth(\tfrac{\hbar \omega}{2 \kB T}). \label{Eq:X-eq01} 
\end{eqnarray} 
%%%
The retarded component of $\check{X}_\bmq(\omega)$ is given by 
$X^R_\bmq(\omega)= S_0^{-1}(\omega-\widetilde{\omega}_\bmq+ \ui \alpha \omega)^{-1}$ 
where $\alpha$ is the Gilbert damping constant, 
and $\widetilde{\omega}_\bmq= \gamma H_0 + \omega_\bmq$ 
is the magnon frequency. 
Here, $\gamma$ is the gyromagnetic ratio and 
$\omega_\bmq= D_{\rm ex} q^2$, 
where $D_{\rm ex}= 2 S_0 J_{\rm ex} a_S^2$ is the spin-wave stiffness constant 
with $J_{\rm ex}$ and $a_S^3$ 
being the exchange energy and the effective block spin volume.

In the nonmagnetic metal, the dynamics of the spin density ${\bm s}$ 
can be described by the action~\cite{Hertz74} 
%%%
\begin{eqnarray}
  {\cal S}_N &=& \int_{C}\; dt \sum_\bmk 
  s^{+}_{-\bmk}(t) 
  [{\chi_{\bmk}( \ui \partial_t)}]^{-1} 
  s^-_{\bm k}(t), 
  \label{Eq:S_N01} 
\end{eqnarray}
%%%
where 
$s^\pm_\bmk = (s^x_\bmk \pm \ui s^y_\bmk)/2$ 
is defined by ${\bm s}_\bmk = {N_N}^{-1/2} 
\sum_{\bmp} c^\dag_{\bmp+\bmk} {\bm \sigma} c_{\bmp}$ 
with ${\bm \sigma}$, 
$c^\dag_\bmp=(c^\dag_{\bmp,\uparrow},c^\dag_{\bmp,\downarrow})$, 
and $N_N$ 
being the Pauli matrices, 
the electron creation operator for spin projection 
$\uparrow$ and $\downarrow$, and the number of atoms 
in the nonmagnetic metal.  
The equilibrium spin-density propagator is given by 
%%%
\begin{eqnarray}
\check{\chi}_{\bmk}(\omega) &=& 
\left( { \chi^R_\bmk (\omega), \atop 0,} 
     { \chi^K_\bmk (\omega)\atop \chi^A_\bmk (\omega)} \right) 
\end{eqnarray}
%%%
with the following equilibrium condition: 
%%%
\begin{eqnarray}
\chi^A_\bmk(\omega)= [\chi^R_\bmk(\omega)]^*,&&
\chi^K_\bmk(\omega)= 2 \ui \, {\rm Im} \chi^R_\bmk(\omega)
\coth(\tfrac{\hbar \omega}{2 \kB T} ). \label{Eq:chi-eq01} 
\end{eqnarray}
%%%
The retarded part of $\check{\chi}$ is given by~\cite{Fulde68} 
$\chi^R_\bmk(\omega)= \chi_N 
(1+ \lambda_N^2 k^2- \ui \omega \tau_{\rm sf})^{-1}$ 
with $\chi_N$, $\lambda_N$, and $\tau_{\rm sf}$ being 
the paramagnetic susceptibility, spin diffusion length, 
and spin relaxation time, the form of which is consistent 
with the corresponding diffusive Bloch equation [see Eq.~(\ref{Eq:Bloch01}) below]. 

Finally, the interaction between magnons and spin density at the interface 
is given by 
%%%
\begin{eqnarray}
  {\cal S}_{F\mathchar`-N} &=&   \int_{C} dt \; 
  \sum_{{\bm k},{\bm q}} 
  \frac{S_0 {\cal J}^{\bmk-\bmq}_{\rm sd}}{\sqrt{N_F N_N}} 
  \; {\bm m}_{-\bmq}(t) \cdot {\bm s}_{\bmk}(t), 
  \label{Eq:S_int'}
\end{eqnarray}  
%%%
where ${\cal J}^{\bmk-\bmq}_{\rm sd}$ is the Fourier 
transform of ${\cal J}_{\rm sd} (\bmr) = J_{\rm sd} \xi_0(\bmr)$ 
with $J_{\rm sd}$ being the $s$-$d$ exchange interaction 
between conduction-electron spins and localized spins, and 
$\xi_0({\bm r})= \sum_{{\bm r}_0 \in {\rm N\mathchar`-N \, interface}} 
a_S^3 \delta({\bm r}-{\bm r}_0)$. 

%%%%%%%%%%%%%%%%%%%%%%%%%%%%%%%%%%%%%
%%%%%%%%%%%%%%%%%%%%%%%%%%%%%%%%%%%%%

It is instructive to point out that in the spin-wave region 
and in the classical limit 
with negligible quantum fluctuations, 
a system described by %the actions 
Eqs.~(\ref{Eq:S_F01}), (\ref{Eq:S_N01}), and (\ref{Eq:S_int'}) 
is equivalent~\cite{Dominicis75,Schmid82} 
to a system described by the stochastic 
Landau-Lifshitz-Gilbert equation, 
%%%
\begin{eqnarray}
  \partial_t {\bm M} &=& 
  [
  \gamma ( {\bm H}_{\rm eff}+ {\bm h} ) 
  -\tfrac{{\cal J}_{\rm sd}}{\hbar} {\bm s}
  ] \times 
  {\bm M}
  + \tfrac{\alpha}{M_s} {\bm M} \times \partial_t {\bm M}, 
  \label{Eq:LLG01} 
\end{eqnarray}
%%%
coupled with the Bloch equation,~\cite{Zhang04} 
%%%
\begin{eqnarray}
 \partial_t {\bm s}&=& (D_N {\bm \nabla}^2 - \tau_{\rm sf}^{-1}) 
 \delta {\bm s} 
 +\tfrac{{\cal J}_{\rm sd}}{\hbar M_s} {\bm M} \times 
     {\bm s} + {\bm l}, 
     \label{Eq:Bloch01}
\end{eqnarray}
%%%
where ${\bm H}_{\rm eff}= H_0 \hat{\bm z}+ (D_{\rm ex}/\gamma)\nabla^2 ({\bm M}/M_s)$, 
$D_N= \lambda_N^2/\tau_{\rm sf}$ is the diffusion constant, and 
$\delta {\bm s}(\bmr) = {\bm s}(\bmr)-s_0 \xi_0(\bmr){\bm M}(\bmr)/M_s$ 
is the spin accumulation with the local equilibrium spin density 
$s_0=\chi_N S_0 J_{\rm sd}/\hbar$. 
The noise field ${\bm h}$ represents 
thermal fluctuations in $F$ 
with $\langle h_i(\bmr,t) \rangle = 0$ 
and 
$\langle h_i(\bmr,t) h_j (\bmr',t')\rangle = 
\frac{2 k_B T({\bm r}) \alpha}{\gamma M_s} 
\delta_{ij} \delta(\bmr-\bmr') \delta(t-t')$,~\cite{Brown63} 
while the noise source ${\bm l}$ in $N$ 
satisfies $\langle l_i(\bmr,t) \rangle = 0$ and 
$\langle l_i(\bmr,t) l_j (\bmr',t')\rangle = 
\frac{2 k_B T({\bm r}) \chi_N a^3}{\tau_{\rm sf}} 
\delta_{ij} \delta(\bmr-\bmr') \delta(t-t')$~\cite{Ma75} 
with the lattice constant $a$, 
both of which are postulated by the fluctuation-dissipation theorem.

In this section we focus on the ``local'' spin injection from $F_1$ into $N_1$. 
The spin current induced in $N_1$ can be calculated 
from the linear response expression of the 
magnon-mediated spin injection given in the Appendix~\ref{Ch:Linear-response} 
[Eq.~(\ref{Eq:I_s01})]. 
Consider the process $P_1$ shown in Fig.~\ref{Fig:fig3_SSEins} (a) 
where magnons travel around the ferromagnet $F_1$ without feeling 
the temperature difference between $F_1$ and $F_2$. 
Using the standard rules of constructing the 
Feynman diagram in Keldysh space,~\cite{Rammer86} 
the corresponding interface Green's function 
$\check{C}_{\bmk,\bmq}(\omega)$ for the correlation 
between the magnons in $F_1$ and the spin density in $N_1$ 
[Eq.~(\ref{Eq:I_s01})] can be written in the form 
%%%
\begin{eqnarray}
  \check{C}_{\bmk,\bmq} (\omega) &=& 
  \frac{{\cal J}^{\bmk-\bmq}_{\rm sd} {S_0} }{\sqrt{N_N N_F}}
  \check{\chi}_\bmk(\omega) \check{X}_\bmq(\omega), 
  \label{Eq:C-func01}
\end{eqnarray}
%%%
where $N_N$ and $N_F$ are the number of lattice sites 
in $N_1$ and $F_1$. 
Substituting Eq.~(\ref{Eq:C-func01}) into Eq.~(\ref{Eq:I_s01}) 
and employing the equilibrium conditions 
[Eqs.~(\ref{Eq:X-eq01}) and ~(\ref{Eq:chi-eq01})], 
we obtain the expression for the injected spin current 
%%% 
\begin{eqnarray}
  I_s^{N_1} &=& 
  -\frac{4 N_{\rm int} {J}_{\rm sd}^2S_0^2 }{\sqrt{2}\hbar^2 N_{N} N_{F}} 
   \sum_{\bmq,\bmk} 
   \int_{\omega} 
    {\rm Im} \chi_{\bmk}^{R}(\omega) 
    {\rm Im} X_{\bmq}^{R}(\omega)  \nonumber \\
    && \quad \times 
      \left[ \coth(\tfrac{\hbar \omega}{2 \kB T_{{N_1}}}) 
         - \coth(\tfrac{\hbar \omega}{2 \kB T_{{F_1}}})   \right], 
      \label{Eq:Is_local01} 
\end{eqnarray}
%%%
where we have introduced the shorthand notation 
$\int_\omega = \int_{-\infty}^{\infty} \frac{d \omega}{2 \pi}$, 
and $N_{\rm int}$ is the number of localized spins at the $N_1$-$F_1$ interface 
playing a role of the number of channels. 
The $\omega$ integration can be performed by picking up 
only magnon poles under the condition 
$\alpha \hbar \widetilde{\omega}_q  \ll \kB T_{N_1}, \kB T_{F_1}$ (always satisfied 
for YIG), giving 
$\int_\omega {\rm Im} \chi_\bmk(\omega){\rm Im} X_\bmq(\omega) 
[\coth(\frac{ \hbar \omega}{2 \kB T})] 
\approx -\frac{1}{2}{\rm Im}\chi_\bmk( \widetilde{\omega}_q)
[\coth(\frac{ \hbar \widetilde{\omega}_\bmq}{2 \kB T})] $. 
By making the classical approximation 
$\coth(\tfrac{\hbar \widetilde{\omega}_\bmq}{2 \kB T}) 
\approx \tfrac{2 \kB T}{ \hbar \widetilde{\omega}_\bmq}$, 
we obtain 
%%%
\begin{eqnarray}
  I_s^{N_1} &=& 
  \frac{N_{\rm int} J_{\rm sd}^{2} S_0 \chi_N \tau_{\rm sf}} 
       {2 \sqrt{2} \pi^4 \hbar^3 (\lambda_N/a)^3 } 
  \Upsilon_1 
  \kB (T_{{N_1}}-T_{{F_1}}), \label{Eq:Is_local02} 
\end{eqnarray}
%%%
where 
$\Upsilon_1 = 
\int_0^1 d x \int_0^1 d y 
\frac{x^2 \sqrt{y}}{ 
[(1+x^2)^2+ y^2 (2 J_{\rm ex} S_0 \tau_{\rm sf}/\hbar)^2  ]}
$ with the dimensionless variables 
$x= {\bm k} \lambda_N$ 
and $y= \hbar \omega_q/(2 J_{\rm ex} S_0)$, 
and we used the relation 
$N_F^{-1}\sum_\bmq = (2 \pi)^{-2} \int \sqrt{y} dy$. 

\section{Magnon-mediated spin Seebeck effect \label{Sec:nonlocal}}

\begin{figure}[t]  
  \begin{center}
    \scalebox{0.65}[0.65]{\includegraphics{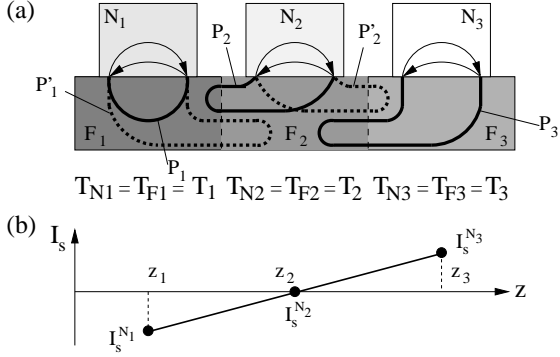}}
  \end{center}
\caption{(a) Feynman diagrams expressing the spin current 
injected from the ferromagnet ($F$) to the nonmagnetic metals ($N$). 
The thin solid lines with arrows (bold lines without arrows) 
represent electron propagators (magnon propagators). 
(b) Spatial profile of the calculated spin current. }
\label{Fig:fig3_SSEins}
\end{figure}

Equation~(\ref{Eq:Is_local02}) means that, through the ``local'' process 
$P_1$ shown in Fig.~\ref{Fig:fig3_SSEins}(a), 
the spin current is {\it not} injected 
into the nonmagnetic metal $N_1$ 
when $F_1$ and $N_1$ have the same temperature. 
That is, the ``local'' process cannot 
explain the experiment~\cite{Uchida10} where 
no temperature difference exists between the YIG film 
and the attached Pt film. 
A way to account for the experiment within the ``local'' picture 
is to invoke a difference between the phonon temperature and magnon 
temperature.~\cite{Xiao10} 
In this paper, on the other hand, we take a different route 
and consider the effect of temperature gradient {\it within} the YIG film 
on the spin injection into the Pt terminal. 

The basic idea of our approach is as follows. 
The above result [Eq.~(\ref{Eq:Is_local02})] that the injected spin current 
vanishes when $T_{F_1}=T_{N_1}$ originates from the equilibrium condition 
of the magnon propagator [Eq.~(\ref{Eq:X-eq01})]. When magnons deviate from 
local thermal equilibrium by allowing the magnons to feel the temperature 
gradient inside the ferromagnet, the magnon propagator cannot be written in 
the equilibrium form, and it generates a nontrivial contribution 
to the thermal spin injection. 
The relevant ``nonlocal'' process $P'_1$ is shown in Fig.~\ref{Fig:fig3_SSEins}(a) 
in which magnons feel the temperature difference between $F_1$ and $F_2$. 
The interaction between $F_1$ and $F_2$ is described by the action 
%%%
\begin{eqnarray}
  {\cal S}_{F\mathchar`-F} &=& \int_{C} dt   \;
  \sum_{\bmq,\bmq'} 
  \frac{2 {\cal J}^{\bmq-\bmq'}_{\rm ex} S_0^2}{N_F} 
  \; {\bm m}_\bmq(t) \cdot {\bm m}_{-\bmq'}(t), 
  \label{Eq:S_int}
\end{eqnarray}  
%%%
where ${\cal J}^{\bmq-\bmq'}_{\rm ex}$ is the Fourier 
transform of ${\cal J}_{\rm ex} (\bmr) = J_{\rm ex} \xi_1(\bmr)$ 
with $\xi_1({\bm r})= \sum_{{\bm r}_0 \in {\rm F\mathchar`-F \;interface}} 
a_S^3 \delta({\bm r}-{\bm r}_0)$. 

%%%%%%%%%%%%%%%%%%%%%%%%%%%%%%%%%%%%%
%%%%%%%%%%%%%%%%%%%%%%%%%%%%%%%%%%%%%

We now regard the whole of the magnon lines 
appearing in the process $P'_1$ as a single magnon propagator 
$\delta \check{X}_\bmq(\omega)$, namely, 
%%%
\begin{eqnarray}
\delta \check{X}_\bmq(\omega) &=& 
\frac{1}{N_F^2}\sum_{\bmq'} |{\cal J}_{\rm ex}^{\bmq-\bmq'}|^2 
\check{X}_\bmq (\omega) \check{X}_{\bmq'} (\omega) \check{X}_\bmq (\omega). 
\end{eqnarray}
%%%
Then the propagator is decomposed into the local-equilibrium part 
and nonequilibrium part as~\cite{Michaeli09} 
%%% 
\begin{eqnarray} 
  \delta \check{X}_\bmq (\omega) &=& 
  \delta \check{X}^{l \mathchar`-eq}_\bmq (\omega) 
  + \delta \check{X}^{n \mathchar`-eq}_\bmq (\omega), 
  \label{Eq:dX_noneq01} 
\end{eqnarray}
%%%
where 
%%%
\begin{equation}
\delta \check{X}^{l \mathchar`-eq}_\bmq = 
\left({\delta {X}^{l \mathchar`-eq,R}_\bmq, \atop 0,} 
{ {\delta X^{l \mathchar`-eq,K}_\bmq } 
\atop \delta {X}^{l \mathchar`-eq, A}_\bmq} \right) 
\label{Eq:dX_loceq01}
\end{equation}
%%%
is the local-equilibrium propagator satisfying the 
local-equilibrium condition, i.e., 
$\delta X^{l \mathchar`-eq,A}_\bmq = 
[\delta X^{l \mathchar`-eq,R}_\bmq]^* $
and $\delta X^{l \mathchar`-eq,K}_\bmq= 
[\delta X^{l \mathchar`-eq,R}_\bmq- \delta X^{l \mathchar`-eq,A}_\bmq] 
\coth(\tfrac{\hbar \omega}{2 \kB T} ) $ 
with 
%%%
\begin{eqnarray}
  \delta X^{l \mathchar`-eq,R}_\bmq(\omega) &=& 
  \frac{1}{N_F^2} \sum_{\bmq'} |{\cal J}_{\rm ex}^{\bmq-\bmq'}|^2
  \Big( X^R_\bmq(\omega) \Big)^2 X^R_{\bmq'}(\omega), 
\end{eqnarray}
%%%
while 
%%%
\begin{equation}
  \delta \check{X}^{n \mathchar`-eq}_\bmq = 
  \left( {0, \atop 0,} 
       { {\delta X^{n \mathchar`-eq,K}_\bmq } 
         \atop 0} \right) 
\end{equation} 
%%%
is the nonequilibrium propagator 
with $\delta {X}_\bmq^{n \mathchar`-eq,K} (\omega)$ given by 
%%%
\begin{eqnarray}
  \delta {X}_\bmq^{n \mathchar`-eq,K} (\omega) &=& 
  \sum_{\bmq'}  \frac{|2{\cal J}_{\rm ex}^{\bmq-\bmq'}S_0|^2}{N_F^2} 
  \Big[ {X}^R_{\bmq'}(\omega)-{X}^A_{\bmq'}(\omega) \Big] \nonumber \\ 
    &\times & 
  |{X}^R_\bmq(\omega)|^2  
  \big[ \coth(\tfrac{\hbar \omega}{2 \kB T_{F_2}}) 
    - \coth(\tfrac{\hbar \omega}{2 \kB T_{{F_1}}}) \big] . 
\end{eqnarray}
%%%
Note that the local equilibrium propagator [Eq.~(\ref{Eq:dX_loceq01})] 
does not contribute to the ``nonlocal'' spin injection.

When we substitute Eq.~(\ref{Eq:dX_noneq01}) into 
Eq.~(\ref{Eq:I_s01}) and use Eq.~(\ref{Eq:C-func01}) with 
$\check{X}_\bmq(\omega)$ being replaced by $\delta \check{X}_\bmq(\omega)$, 
we obtain the following expression for the 
magnon-mediated thermal spin injection: 
%%% 
\begin{eqnarray}
  I_s^{N_1} &=&  \frac{-4J^2_{\rm sd}S_0 (2 J_{\rm ex} S_0)^2 N_{\rm int} N'_{\rm int} }
  {\sqrt{2} \hbar^2 N_F^3 N_N } 
     \sum_{\bmq,\bmq',\bmk} 
     \int_\omega 
              {\rm Im} \chi_{\bmk}^R(\omega) 
              \nonumber \\
     &\times& 
              |X_{\bmq}^{R}(\omega)|^2 
                {\rm Im} X_{\bmq'}^R(\omega)  
                [ \coth(\tfrac{\hbar \omega}{2 \kB T_{1}}) 
                  - \coth(\tfrac{\hbar \omega}{2 \kB T_{2}}) ], 
                \label{Eq:Is_nonlocal01} 
\end{eqnarray}
%%%
where $N'_{\rm int}$ is the number of localized 
spins at the $F_1$-$F_2$ interface, and we used 
$T_{N_i}= T_{F_i}= T_i$ $(i=1,2)$. 
The $\omega$ integration can be performed as before, giving 
$
\int_\omega
{\rm Im} \chi_{\bmk}^R(\omega) 
|X_{\bmq}^R(\omega)|^2 
{\rm Im} X_{\bmq'}^R(\omega)  
[ \coth(\tfrac{\hbar \omega}{2 \kB T_{1}})- \coth(\tfrac{\hbar \omega}{2 \kB T_{2}}) ] 
\approx 
\frac{- \pi }
{2 \alpha \widetilde{\omega}_{\bmq}} 
\delta(\omega_{\bmq}-\omega_{\bmq'}) 
{\rm Im}\chi_{\bmk}^R(\widetilde{\omega}_\bmq) 
[ \coth(\tfrac{\hbar \widetilde{\omega}_\bmq}{2 \kB T_{1}}) 
- \coth(\tfrac{\hbar \widetilde{\omega}_\bmq}{2 \kB T_{2}}) ] 
$, which suggests that the magnon 
modes with different $\bmq$'s do not interfere with each other. 
With the classical approximation 
$\coth(\tfrac{ \hbar \widetilde{\omega}_\bmq}{2 \kB T}) \approx 
\tfrac{2 \kB T}{\hbar \widetilde{\omega}_\bmq}$, we obtain 
%%% 
\begin{eqnarray}
  I_s^{N_1} &=&  
  \frac{N_{\rm int} (J^2_{\rm sd} S_0) \chi_N \tau_{\rm sf} (a/\lambda_N)^3} 
       {8 \sqrt{2} \pi^5 \hbar^3 \alpha (\Lambda/a_S)} 
       \Upsilon_2 \kB \delta T, %(T_1- T_2), 
      \label{Eq:Is_nonlocal03} 
\end{eqnarray}
%%% 
where $\delta T= T_{1}- T_{2}$, 
$\Lambda$ is the size of $F_1$ along the temperature gradient, 
and $\Upsilon_2 = 
\int_0^1 d x \int_0^1 d y 
\frac{y^2} 
{[(1+x^2)^2+ y^2 (2 S_0 J_{\rm ex} \tau_{\rm sf}/\hbar)^2  ]}$ 
which is approximated 
as $\Upsilon_2 \approx 0.1426$ 
($\Upsilon_2 \approx 0.337 \hbar/ 2 S_0 J_{\rm ex} \tau_{\rm sf}$) 
for $2 S_0 J_{\rm ex} \tau_{\rm sf}/\hbar \!\alt\! 1$ 
(for $2 S_0 J_{\rm ex}\tau_{\rm sf}/\hbar \gg 1$). 

The spin current $I_s^{N_3} $ injected into the right terminal $N_3$ 
can be calculated in the same manner by considering the process $P_3$, 
which gives $I_s^{N_3}= -I_s^{N_1}$ 
from the relation $T_1-T_2= -(T_3-T_2)$. 
The spin current $I_s^{N_2}$ 
injected into the middle terminal $N_2$ vanishes 
because the two relevant processes ($P_2$ and $P'_2$) cancel out. 
Therefore, we obtain the spatial profile of the injected spin current 
as shown in Fig.~\ref{Fig:fig3_SSEins}(b). 
Note that the effect of the spatial dependence of magnetization 
${\bm M}[T({\bm r})]$ through the local temperature $T({\bm r})$ 
is already taken into account in our treatment because the temperature 
dependence of ${\bm M}$ in the magnon region 
is automatically described by the number of thermal magnons 
discussed in this paper.

For an order of magnitude estimation, 
we compare Eq.~(\ref{Eq:Is_nonlocal03}) with the experiment.~\cite{Uchida10} 
By using $\Theta_H \approx 0.0037$,~\cite{Kimura07,Hoffmann10} 
$\rho = 15.6 \times 10^{-8} \, \Omega  {\rm m} $, 
$w = 0.1 \, {\rm mm}$, 
$\lambda_N \approx 7 \, {\rm nm}$, 
$\tau_{\rm sf} \approx 1 \, {\rm ps}$, 
$a =  2 \, \AA $, 
$a_S =  12.3 \, \AA$, 
$S_0 =16$, 
$\alpha \approx 5 \times 10^{-5}$,~\cite{Kajiwara10} 
$\chi_N = 1 \times 10^{-6} \, {\rm cm}^3/{\rm g}$,~\cite{Kriessman54} 
and $N_{\rm int}= 0.1 \times 4 {\rm mm}^2/a_S^2$, 
the $s$-$d$ exchange coupling extracted from the previous ferromagnetic resonance 
experiment~\cite{Kajiwara10} ($J_{\rm sd} \approx 10 \, {\rm meV}$) 
can account for the spin Seebeck voltage 
$V_{\rm ISHE}/\delta T \approx 0.1 \, \mu{\rm V}/{\rm K}$ 
observed at room temperature. 

Finally, we comment on the issue of length scales associated with the SSE. 
In the original SSE experiment for a {\it metallic} ferromagnet,~\cite{Uchida08} 
the signal maintained over several millimeters was a big surprise because 
the spin diffusion length for that system is much shorter than a millimeter. 
Concerning the magnon-mediated SSE in an {\it insulating} 
magnet~\cite{Uchida10} which we have discussed, 
it is of crucial importance to recognize that the length scale relevant to 
the SSE is related to magnon density fluctuations and is given by 
{\it longitudinal} fluctuations of magnons, 
while the magnon mean free path is related to magnon dephasing 
and is given by {\it transverse} fluctuations of magnons.~\cite{com01} 
It was shown by Mori and Kawasaki~\cite{Mori62} 
that these two length scales do not coincide with each other since 
they obey quite different dynamics, and it was demonstrated that 
in a certain situation the length scale of magnon density fluctuations 
(which is relevant to the SSE as well) is much longer than the magnon mean free path 
[see Eq.(6.33) in Ref.~\onlinecite{Mori62} where the length scale of 
long-wavelength magnon density fluctuations is infinitely long].~\cite{com02} 
The notion of these two different length scales is the key to understanding  
the length scales observed in the 
SSE experiment in an insulating magnet.~\cite{Uchida10}

\section{Conclusion \label{Sec:conclusion}}

We have developed a theory of the magnon-mediated 
spin Seebeck effect in terms of the canonical framework of 
describing transport phenomena, 
i.e., the linear-response theory, 
and shown that it provides us with 
a qualitative as well as quantitative understanding of 
the spin Seebeck effect observed in a prototypical magnet, 
yttrium iron garnet.~\cite{Uchida10} 
Because the carriers of spin current in this scenario are magnons, 
we can obtain a bigger signal for a magnetic material 
with a lower magnon damping 
[see Eq.~(\ref{Eq:Is_nonlocal03}) where the injected spin current 
is inversely proportional to the Gilbert damping constant $\alpha$]. 
An advantage of our linear-response formulation is that it can be easily 
generalized to a situation including degrees of freedom other than magnons, 
e.g., phonons and conduction electrons, to describe 
a more complicated process in the case of metallic~\cite{Uchida08} and 
semiconducting systems,~\cite{Jaworski10} 
and a calculation taking account of the effect of nonequilibrium phonons 
will be reported in a future publication.~\cite{Adachi-next} 
A numerical approach to the SSE is also developed in Ref.~\onlinecite{Ohe11}. 
We believe that the present approach stimulates further research 
on the spin Seebeck effect. 

\acknowledgments

We are grateful to E. Saitoh, K. Uchida, G. E. W. Bauer, and J. Ieda 
for helpful discussions. 
This work was financially supported by 
a Grant-in-Aid for Scientific Research on Priority Areas (No. 19048009) 
and a Grant-in-Aid for Young Scientists (No. 22740210) from MEXT, Japan.

\appendix 

\section{Linear-response expression of magnon-induced 
spin injection\label{Ch:Linear-response}} 
The Gaussian action for conduction electrons in the 
nonmagnetic metal $N_i$ ($i=1,2,3$) is given by 
%%%
\begin{eqnarray}
  {\cal S}_{N} &=& 
  \int_{C} dt \; 
  \sum_{\bmp,\bmp' } 
  c^\dag_\bmp (t) \bigg\{ 
  \ui \partial_t - \Big( 
  \epsilon_\bmp \delta_{\bmp,\bmp'} \qquad \qquad \nonumber \\
  && + U_{\bmp-\bmp'} [1+ \ui \eta_{\rm so} 
  {\bm \sigma}\cdot (\bmp \times \bmp')] 
  \Big) \bigg\} c_{\bmp'} (t), 
  \label{Eq:H_el01}
\end{eqnarray}
%%%
where $c^\dag_\bmp=(c^\dag_{\bmp,\uparrow},c^\dag_{\bmp,\downarrow})$ 
is the electron creation operator for spin projection 
$\uparrow$ and $\downarrow$, 
$U_{\bmp-\bmp'}$ 
is the Fourier transform of the impurity potential 
$U_{\rm imp} \sum_{\bmr_0 \in {\rm impurities}} \delta(\bmr-\bmr_0)$, 
and $\eta_{so}$ measures the strength of the 
spin-orbit interaction.~\cite{Takahashi08} 

At the ferromagnet/nonmagnetic-metal interface, the magnetic interaction 
between conduction-electron spin density and localized spin is described 
by the $s$-$d$ interaction [Eq.~(\ref{Eq:S_int'})]. 
The spin current induced in the nonmagnetic metal $N_1$ 
can be calculated as the rate of change of the spin accumulation in $N_1$, 
i.e., 
$I_s^{N_1}(t) \equiv 
\sum_{\bmr \in N_1} \langle \partial_t s^z({\bm r}, t) \rangle 
= \langle \partial_t \widetilde{s}^z_{\bmk_0}(t) \rangle_{\bmk_0 \to {\bm 0}} $, 
where $\langle \cdots \rangle$ means the statistical average at a given time $t$, 
and $\widetilde{\bms}_\bmk= \sqrt{N_N} \bms_\bmk$ 
with ${\bm s}$ being defined below Eq.~(\ref{Eq:S_N01}). 

The Heisenberg equation of motion for $\widetilde{s^z}_{\bmk_0}$ gives 
%%%
\begin{eqnarray} 
  \partial_t \widetilde{s}^z_{\bmk_0} 
  &=& \sum_{\bmq,\bmk} 
  \frac{\ui {\cal J}^{\bmk- \bmq}_{\rm sd} S_0}{\sqrt{2 N_F N_N} \hbar } 
  \Big( m^+_{-\bmq} [s^-_{\bmk},s^z_{\bmk_0}] 
  + m^-_{-\bmq} [s^+_{\bmk},s^z_{\bmk_0}] \Big) \nonumber \\
  &=& 
  \ui \sum_{\bmq,\bmk} \frac{2{\cal J}^{\bmk-\bmq}_{\rm sd} {S_0} } 
      {\sqrt{2 N_F N_N} \hbar } 
      \Big( m^+_{-\bmq} s^-_{\bmk+\bmk_0} 
      - m^-_{-\bmq} s^+_{\bmk+\bmk_0} \Big), 
\end{eqnarray} 
%%%
where we have used the relation 
$[\widetilde{s}^z_{\bmk},\widetilde{s}^\pm_{\bmk'}] 
= \pm 2 \widetilde{s}^\pm_{\bmk+\bmk'} $, and 
neglected a small correction term arising from the 
spin-orbit interaction assuming that 
the spin-orbit interaction is weak enough 
at the neighborhoods of the interface. 
Then, the statistical average of the above quantity gives 
the following spin current: 
%%%
\begin{eqnarray}
  I^{N_1}_s(t) 
  &=& \sum_{\bmq,\bmk} 
  \frac{-4{\cal J}^{\bmk-\bmq}_{\rm sd} {S_0} } 
  {\sqrt{2 N_F N_N} \hbar} 
  {\rm Re} C^{<}_{\bmk,\bmq}(t,t), 
  \label{Eq:I_s00} 
\end{eqnarray}
%%%
where 
$C^{<}_{\bmk,\bmq}(t,t') = 
- \ui \langle  m^+_{-\bmq}(t') s^-_\bmk(t) \rangle $ 
is the interface Green's function. 
In the steady state, the Green's function $C^{<}_{\bmk,\bmq}(t,t') $ 
depends only on the time difference $t-t'$ as 
$C^{<}_{\bmk,\bmq}(t-t') = 
\int_{-\infty}^\infty \frac{d \omega}{2 \pi} 
{C}^{<}_{\bmq,\bmk}(\omega) e^{- \ui \omega (t-t')}$. 
Adopting the representation~\cite{Larkin75} 
$\check{C} 
= \left({{C^{R}, C^{K}} \atop {0 \;\;\;  ,C^{A}}} \right)$ 
and using $C^{<}= \frac{1}{2} [C^{K}- C^{R} + C^{A}]$, 
we finally obtain 
%%%
\begin{eqnarray}
  I^{N_1}_s &=&   \sum_{\bmq,\bmk} 
  \frac{-2{\cal J}^{\bmk-\bmq}_{\rm sd} {S_0} }
  {\sqrt{2 N_F N_N} \hbar} 
  \int_{-\infty}^\infty \frac{d \omega}{2 \pi} 
  {\rm Re} C^{K}_{\bmk,\bmq}(\omega) 
  \label{Eq:I_s01}
\end{eqnarray}
%%%
for the spin current $I^{N_1}_s$ in a steady state. 
As in the case of tunneling charge current driven by 
a voltage difference,~\cite{Caroli71} the spin current $I_s^{N_1}$ 
can be calculated systematically.


\begin{thebibliography}{99}
\bibitem{com01}
  Note that our scenario does not rely on the long propagation length 
  of dipole magnons ($\sim$ several millimeters) observed in 
  Ref.~\onlinecite{Schneider08}, 
  since the magnons which we have discussed here are of exchange origin.
\bibitem{com02}
  An analogous situation occurs for conduction electrons in a disordered 
  metal~\cite{Vollhardt80} where the length scale associated with the 
  long-wavelength density fluctuations of conduction electrons is infinite 
  due to the charge conservation while that related to the dephasing 
  of conduction electrons, i.e., the mean free path of conduction electrons, 
  is as short as several nanometers.
\end{thebibliography}

\end{document}